\documentclass[12pt]{article}
\usepackage[xdvi]{graphicx} 
\usepackage{epsfig}
\usepackage{amsfonts,amssymb,amsmath}
\usepackage{subfigure}

\newcommand{\ga}{\ensuremath{\gamma}}

\newcommand{\mn}{\ensuremath{{\mu\nu}}}

\newcommand{\eps}{\ensuremath{\epsilon}}
\newcommand{\s}{\ensuremath{\sigma}}

\newcommand{\del}{\ensuremath{\partial}}

\newcommand{\pl}{\ensuremath{{\textrm{pl}}}}
\newcommand{\half}{\frac{1}{2}}

\newcommand{\be}{\begin{equation}}
\newcommand{\ee}{\end{equation}}
\newcommand{\m}{\mathcal}
\newcommand{\ba}{\begin{eqnarray}}
\newcommand{\ea}{\end{eqnarray}}

\begin{document}

%\rightline{OUTP-04/18P}

%%                      Title here 
%%
%\draft
%\title{A short review of ``DGP Specteroscopy".}
%\author{Antonio Padilla}
%\address{School of Physics and Astronomy, University Pary, University of Nottingham, Nottingham NG7 , Spain \\~\\Email:
 %   padilla@ffn.ub.es}

\begin{center}
{\Large \bf A short review of ``DGP Specteroscopy".}
\end{center}
\vskip 1cm 
\renewcommand{\thefootnote}{\fnsymbol{footnote}}

\centerline{\bf Antonio Padilla\footnote{antonio.padilla@nottingham.ac.uk}}
\vskip .5cm

\centerline{\it School of Physics and Astronomy}
\centerline{\it University Park, University of Nottingham, Nottingham NG7 2RD, UK}

\setcounter{footnote}{0} \renewcommand{\thefootnote}{\arabic{footnote}}
 
%%                      Text starts here 
%% 

\begin{abstract}
In this paper we provide a short review of the main results developed in hep-th/0604086. We focus on linearised vacuum perturbations about the self-accelerating branch of solutions in the DGP model. These are shown to contain a ghost in the spectrum for any value of the brane tension. We also comment on hep-th/0607099, where some counter arguments have been presented.
\end{abstract}

Recent observations of high redshift supernovae suggest that dark energy accounts for roughly  $70 \%$ of the energy content of our universe~\cite{Riess}. This dark energy is consistent with a small positive cosmological constant, $\Lambda \sim 10^{-12}$ (eV)${}^4$, exerting negative pressure on the universe, causing its expansion to accelerate. If we wish to resort to effective field theory methods to explain the origin of the cosmological constant, we typically run into an horrendous fine tuning problem. For a field theory cut off at the Planck scale, $m_\pl$, the natural value of the cosmological constant would be of the order $m_\pl^4$, which is  $10^{120}$ orders of magnitude larger than the observed value.

This problem has inspired a search for alternative explanations of the current cosmic acceleration. One possibility is that it is due to new gravitational physics kicking in at the current Hubble scale, $H \sim 10^{-34}$eV (see, for example~\cite{accn}). In this paper we will focus on the DGP model~\cite{DGP}, which has arguably received more attention than any other model in which gravity is modified on ultra large scales. The model consists of a $\mathbb{Z}_2$ symmetric $3$-brane embedded in $5 D$ Minkowski space, described by the following action
\be
%\fl 
S=2M_5^3\int_\textrm{bulk} \sqrt{-g} R+4M_5^3\int_\textrm{brane} \sqrt{-\gamma} K+\int_\textrm{brane} \sqrt{-\gamma}\left(M_4^2 \mathcal{R}-\sigma+\mathcal{L}_\textrm{matter}\right) \label{action}
\ee
where $g_{ab}$ is the bulk metric with corresponding Ricci tensor $R_{ab}$. The brane has induced metric $\gamma_\mn$ with corresponding Ricci tensor $\mathcal{R}_\mn$,  and extrinsic curvature $K_\mn$. The key feature here is the intrinsic curvature induced on the brane by matter loop corrections~\cite{loops}, or finite width effects~\cite{width}. Note that we have included an explicit brane tension $\sigma$, and additional matter lagrangian, $\m{L}_\textrm{matter}$.
The governing equations of motion in the bulk are simply the vacuum Einstein equations
\be
G_{ab}=R_{ab}-\half R g_{ab}=0
\ee
whereas the boundary conditions at the brane are given by the Israel junction conditions
\be
%\fl \qquad
\Theta_\mn=2M_5^3(K_\mn-K \ga_\mn)+M_4^2 \left(\m{R}_\mn-\half \m{R}\ga_\mn \right)+\frac{\sigma}{2}\ga_\mn= \frac{1}{2}T_\mn\label{israel}
\ee
where $T_\mn=-\frac{2}{\sqrt{-\ga}}\frac{\del(\sqrt{-\ga} \m{L}_\textrm{matter})}{\del \ga^\mn}$. In the absence of any additional matter, we can set $T_\mn=0$, and derive the following background spacetimes
\be
ds^2=\bar{g}_{ab} dx^adx^b =e^{2\eps H |y|}\left(dy^2+\bar
\gamma_{\mu\nu}dx^\mu dx^\nu\right)  \label{background}
\end{equation}
where $\eps=\pm 1$, and 
\be 
\bar \gamma_{\mu\nu}dx^\mu dx^\nu=-dt^2+ e^{2 Ht} \, d
\vec x^2 \label{warpds} 
\ee
The bulk spacetime  corresponds to $-\infty<y<0$ and $0<y
<\infty$, with a de Sitter brane positioned at $y=0$. The brane can be thought of as a $4D$ hyperboloid embedded in a  $5D$ Minkowski bulk. The sign of $\eps$ determines whether the bulk  spacetime corresponds to  two copies of  the exterior of the hyperboloid ($\eps=+1$), or two copies of the interior ($\eps=-1$). The solution
with $\eps=-1$ is commonly referred to as the {\it normal} branch
whereas  the solution with $\eps=+1$ is referred to as the {\it
self-accelerating} branch, a terminology which will become
transparent shortly. Note that the  metric in (\ref{warpds}) represents
the $4D$ de Sitter geometry in spatially flat coordinates, which covers only one half of the de Sitter hyperboloid.

The value of the intrinsic curvature on the brane can be related to the brane tension using the Israel equations (\ref{israel}). It turns out that
\be
H=\half H_0 \left(\eps+\sqrt{1+\frac{\s}{3M_5^3H_0}}\right)
\ee
where $H_0=2M_5^3/M_4^2$ is taken  to be the current Hubble scale.  Note that even for vanishing tension, the self accelerating solution gives rise to a de Sitter brane universe with $H=H_0$. The modification of gravity at large distances has enabled us to describe an accelerating universe in the absense of any vacuum energy whatsoever! In contrast, the normal branch  gives rise to a Minkowski brane as $\s \to 0$, and is of less interest  phenomenologically.

In ``DGP Specteroscopy"~\cite{spec}, we discussed the stability of linearised perturbations about the background solution (\ref{background}). On the normal branch, these perturbations are well behaved. In contrast, on the self accelerating branch, one is generically haunted by ghosts. In this letter, we will review the discussion of  linearised perturbations about the  self accelerating solution. For brevity, we will restrict attention to $\mathbb{Z}_2$ symmetric fluctuations about the vacuum ($T_\mn=0$). A more complete discussion including asymmetric fluctuations and the contribution from additional matter ($T_\mn\neq 0$) can be found in~\cite{spec}.

 Recall that the self-accelerating background solution is given by the metric (\ref{background}) with $\eps=+1$ and the brane positioned at $y=0$. A generic perturbation can be described by $g_{ab}=\bar g_{ab}+\delta g_{ab}$ with the brane position shifted to $y=F(x)$. It is convenient to work in a Gaussian Normal (GN) gauge so that
\be
\delta g_{yy}=\delta g_{\mu y}=0, \qquad \delta g_\mn=e^{H|y|/2} h_\mn(x, y)
\ee 
The tensor $h_\mn$ can be decomposed in terms of the irreducible representations of the $4D$ de Sitter diffeomorpism group
\be
h_{\mu\nu} = h^{\tt TT}_{\mu\nu} + D_\mu A_\nu + D_\nu A_\mu +
D_\mu D_\nu \phi - \frac14 \bar \gamma_{\mu\nu} D^2 \phi +
\frac{h}{4} \bar \gamma_{\mu\nu}  \label{decomps} 
\ee
where $D_\mu$ is the covariant derivative for the $4D$ de Sitter   metric $\bar \ga_\mn$.  The transverse-tracefree tensor  $h^{\tt TT}_{\mu\nu}$ satisfies 
$D^\mu h^{{\tt TT} }{}_\mn = h^{{\tt TT}~\mu}{}_{\mu} = 0$, and has $5$ independent components.  $A_\mu$ is a Lorentz-gauge vector, $D^\mu
A_\mu = 0$, with $3$ independent components, and $\phi$ and $h = h^\mu{}_\mu$
are two scalar fields\footnote{Note that the total number of independent components  correctly adds up  10}.

We can fix the position of the brane to be at $y=0$ whilst remaining in GN gauge, by making the following gauge transformation
\be
y \to y-Fe^{-H|y|}, \qquad x^\mu \to x^\mu-\frac{e^{-H|y|}}{H} D^\mu F
\ee
Although the brane position is now fixed at $y=0$,  the original brane position $F(x)$ still enters the dynamics through a bookkeeping term $h_\mn^{(F)}$ that  modifies the metric perturbation
\be
%\fl \qquad 
h_\mn \to  h^{\tt TT}_{\mu\nu} + D_\mu A_\nu + D_\nu A_\mu +
D_\mu D_\nu \phi - \frac14 \bar \gamma_{\mu\nu} D^2 \phi +
\frac{h}{4} \bar \gamma_{\mu\nu}+h_\mn^{(F)}
\ee
The bookkeeping term is of course pure gauge in the bulk, and is  given by
\be
h_\mn^{(F)}=\frac{2}{H}e^{H|y|/2}\left(D_\mu D_\nu+H^2\bar \ga_\mn\right)F
\ee
We can now substitute our modified expression for $h_\mn$ into the linearised fields equations in the bulk, $\delta G_{ab}=0$, and on the brane, $\delta \Theta_\mn=0$. It turns out that the Lorentz-gauge vector $A_\mu$ is a free field in the linearsed theory and can be set to zero. In addition, the $yy$ and $y\mu$ equations in the bulk imply that one can consistently choose a gauge for which
\be
h=0, \qquad (D^2+4H^2)\phi=0
\ee
Note that we now have $h_\mn= h^{\tt TT}_{\mu\nu}+h_\mn^{(\phi)}+h_\mn^{(F)}$, where the contribution from $\phi(x, y)$ has been  rewritten as follows
\be
h_\mn^{(\phi)}=\left(D_\mu D_\nu+H^2\bar \ga_\mn\right)\phi(x, y)
\ee
This mode is now entirely transverse-tracefree in its own right. In the absence of any additional matter on the brane ($T_\mn=0$), the same is true of the bookkeeping mode, $h_\mn^{(F)}$. This is because the trace of the Israel equation now implies that 
\be(D^2+4H^2)F=0\ee
The entire perturbation $h_\mn(x, y)$ is now completely transverse-tracefree. This greatly simplifies the bulk and brane equations of motion, giving
\begin{eqnarray}
\left[ D^2-2H^2+\del_y^2-\frac{9H^2}{4}\right]h_\mn(x, y)=0 &&\qquad \textrm{ for $|y|>0$} \\
\left[M_4^2(D^2-2H^2)+2M_5^3\left(\del_y-\frac{3H}{2}\right)\right]h_\mn=0 && \qquad\textrm{ at $y=0^+$}
\end{eqnarray}
We now separate variables in the tensor and scalar fields as follows
\be
 h^{\tt TT}_{\mu\nu}(x, y)=\sum_{m} u_m(y) \chi^{(m)}_\mn(x), \qquad \phi(x, y)=W(y)\hat \phi(x)
\ee
where $\chi_\mn^{(m)}$ is $4D$ tensor field of mass $m$ satisfying $(D^2-2H^2)\chi_\mn^{(m)}=m^2\chi_\mn^{(m)}$. Note that $\hat \phi$ is a $4D$ tachyonic field satisfying $(D^2+4H^2)\hat \phi=0$. This is a mild instability which is related to the repulsive nature of inflating domain walls.

We shall now focus on the case where the brane tension is non-vanishing ($\sigma \neq 0$). Assuming that the tensor and scalar equations of motion can be treated independently, we find that there is a continuum of normalisable tensor modes with mass $m^2 \geq 9H^2/4$. In addition, there is also a discrete tensor mode with mass
\be
m_d^2=H_0(3H-H_0)
\ee
and normalisable wavefunction $u_{m_d}(y)=\alpha_{m_d}e^{-|y|\sqrt{\frac{9H^2}{4}-m_d^2}}$. Now, for {\it positive} brane tension $\sigma >0$, one can easily check that $0<m_d^2<2H^2$. For massive gravitons propagating in $4D$ de Sitter, it is well known that masses lying in this range result in the graviton containing a helicity-$0$ ghost~\cite{higuchi}. This means that for $\sigma >0$, the lightest tensor mode contains a helicity-$0$ ghost, and so the system is perturbatively unstable. For {\it negative} brane tension, $m_d^2>2H^2$ and there is no helicity-$0$ ghost in the lightest tensor.

Now consider the scalar equations of motion. The first thing to note is that $h_\mn^{(\phi)}$ behaves like a transverse-tracefree mode with mass $m^2_\phi=2H^2$, because $(D^2-2H^2)h_\mn^{(\phi)}=2H^2 h_\mn^{(\phi)}$. Since none of the tensor modes have this mass, they are all orthogonal to $h_\mn^{(\phi)}$. This means it was consistent to assume that the scalar and tensor equations of motion could indeed be treated independently. It turns out that the scalar has a {\it normalisable} wavefunction $W(y)=e^{-H|y|/2}$, and  the $4D$ scalar $\hat \phi$ is sourced by $F$ via the relation
\be
\hat \phi(x)=\alpha F(x), \qquad \alpha=-\left[\frac{2H-H_0}{H(H-H_0)}\right]
\ee 
This is well defined for $\sigma \neq 0$ since then $H \neq H_0$. $h_\mn^{(\phi)}(x, y)$ may now be thought of as a genuine radion mode, measuring the physical motion of the brane with respect to infinity. It does {\it not} decouple even though we only have a single brane. This property is related to the fact that the background warp factor, $e^{2H|y|}$,  {\it grows} as we move deeper into the bulk.

We have already identified the  helicity-$0$ mode of the lightest tensor as a ghost when $\sigma >0$. When $\sigma<0$, a calculation of the $4D$ effective action will reveal the ghost to be the radion. Taking our solution 
\ba
h_\mn(x, y)&=&\alpha_{m_d}e^{-|y|\sqrt{\frac{9H^2}{4}-m_d^2}}\chi_\mn^{(m_d)}(x)+\alpha e^{-H|y|/2}(D_\mu D_\nu+H^2\bar \ga_\mn)F\nonumber\\
&&\qquad +\frac{2}{H}e^{H|y|/2}(D_\mu D_\nu+H^2\bar \ga_\mn)F+\textrm{continuum modes} \label{soln}
\ea
and inserting it into the action (\ref{action}), we can integrate out the extra dimension. This is made possible by restricting attention to normalisable modes. The result is $S_\textrm{eff}= \int d^4 x \sqrt{-\bar \ga} \m{L}_\textrm{eff}$, where
\be
\m{L}_\textrm{eff}=\m{L}_\textrm{PF}\left[\chi^{(m_d)}\right]-\frac{m_d^2}{4}\chi^{(m_d)~\mn}\left(\chi_\mn^{(m_d)}-\chi^{(m_d)}\bar \ga_\mn\right)-3M_5^3H^2\alpha F(D^2+4H^2) F+\ldots
\ee
$\m{L}_\textrm{PF}$ is the standard Pauli-Fierz lagrangian and ``$\ldots$" denotes the contribution from the continuum of tensor modes. When $\sigma<0$, it turns out that $\alpha>0$, and so we confirm that negative brane tension gives rise to a radion ghost.

Given that there is always a ghost for non-zero tension, one might expect by continuity that this remains the case when $\sigma=0$. To study this more closely, let us first ask whether we can trust the above solutions in the limit where $\sigma \to 0$.  In this limit $H\to H_0$, and the quantity $\alpha$ becomes ill-defined! To understand what has gone wrong, note that the mass of the lightest tensor $m_d^2 \to 2H^2$. This means that it is no longer orthogonal to the radion, $h_\mn^{(\phi)}$, and so we cannot treat the tensor and scalar equations of motion independently. This behaviour can be traced back to  an additional symmetry that appears in the linearised theory in the limit of vanishing brane tension. It is analagous to the ``partially massless limit" in the theory of a massive graviton propagating in de Sitter space~\cite{higuchi}. In {\it that} theory, the equations of motion are invariant under the following redefinition of the graviton field
\be
\chi_\mn^{(\sqrt{2} H)}(x)\to \chi_\mn^{(\sqrt{2} H)}(x)+(D_\mu D_\nu+H^2\bar \ga_\mn)\psi(x)
\ee
This field redefinition has the effect of extracting out part of the helicity-$0$ mode from $\chi_\mn^{(\sqrt{2} H)}$, and as a result of the symmetry this mode disappears from the spectrum. In {\it our} case, this shift must be acompanied by a shift in the scalar field $\phi$,  
\be
\phi(x, y)\to \phi(x, y)-\alpha_{\sqrt{2}H} e^{-H|y|/2}\psi(x)=\phi(x, y)-\lim_{\sigma \to 0}\alpha_{m_d} e^{-|y|\sqrt{\frac{9H^2}{4}-m_d^2}}\psi(x)
\ee
in order to render the overall perturbation, $h_\mn(x, y)$, invariant.  These $\psi$ shifts can be understood as extracting part of the helicity-$0$ mode from $\chi_\mn^{(\sqrt{2} H)}$ and absorbing it into a renormalisation of $\phi$. The symmetry will have the effect of combining the helicity-$0$ mode and the radion into a single degree of freedom. It is only {\it after} fixing this $\psi$ symmetry that we can treat the scalar and tensor equations of motion independently of one another. We might think of extracting the entire helicity-$0$ mode and absorbing it into $ \phi$, or vice versa. Actually, it will be convenient to make a different gauge choice that enables us to take a smooth limit as $\sigma \to 0$~\cite{koyama}.  We start off with the solution for $\sigma \neq 0$ given by equation (\ref{soln}), and   make the field redefinition
\be
\chi_\mn^{(m_d)}\to \m{H}_\mn=\chi_\mn^{(m_d)}+\left(D_\mu D_\nu+H^2\ga_\mn\right)\left[\frac{\alpha}{\alpha_{m_d}}F\right]
\ee
In the limit as $\sigma \to 0$, this has the effect of extracting out part of the helicity-$0$ mode of $\chi_\mn^{(m_d)}$ and asborbing it into a renormalisation of  $\phi$.
\be
\phi(x, y)\to  \lim_{\sigma\to 0}e^{-H|y|/2}\alpha F-\alpha_{m_d} e^{-|y|\sqrt{\frac{9H^2}{4}-m_d^2}}\left[\frac{\alpha }{\alpha_{m_d}}F\right]=-|y|e^{-H|y|/2}F
\ee
It follows that for vanishing tension
\ba
h_\mn(x, y)&=&e^{-H|y|/2}\left[\alpha_{\sqrt{2}H}\m{H}_\mn-|y|(D_\mu D_\nu+H^2\bar \ga_\mn)F\right]\nonumber\\
&&\qquad +\frac{2}{H}e^{H|y|/2}(D_\mu D_\nu+H^2\bar \ga_\mn)F+\textrm{continuum modes}
\ea
where the $4D$ tensor $\m{H}_\mn$ satisfies
\be
\alpha_{\sqrt{2}H}(D^2-4H^2) \m{H}_\mn=-H(D_\mu D_\nu+H^2\bar \ga_\mn)F
\ee
This equation should be understood as the  helicity-$0$ component of $\m{H}_\mn$ being completely determined by the source $F$. A calculation of the $4D$ effective action in this case, now gives
\be
\m{L}_\textrm{eff}=\m{L}_\textrm{PF}\left[\m{H}\right]-\frac{H^2}{2}\m{H}^{\mn}\left(\m{H}_\mn-\m{H}\bar \ga_\mn\right)+\frac{\sqrt{2}M_5^3}{M_4}\m{H}^\mn (D_\mu D_\nu+H^2\bar \ga_\mn)F+\ldots \label{0action}
\ee
A derivation of the Hamiltonian for this action reveals the presence of a scalar degree of freedom whose energy is unbounded from below~\cite{koyama}. This ghost is a combination of the radion and helicity-$0$ mode, and represents the residual scalar degree of freedom left over after fixing the aforementioned $\psi$ symmetry.

We conclude that for any value of the brane tension, perturbations about the self-accelerating branch of DGP contain a ghost. We would like to emphasise that this ghost-like instability is ultimately {\it classical}, and one cannot hide behind a UV completion of DGP to save the day. The ghost couples to matter, and even in the absence of matter we would expect it to couple  to the tensor modes through higher order interactions. Given that its energy is unbounded from below, the ghost will continually dump its energy into the other fields, rapidly destroying the self-accelerating solution. We might expect the rate of  the instability to go like the frequency of oscillation of the coupled fields. Given that we have an entire tower of heavy tensor modes, this frequency could be very large indeed.

We would like to end our discussion with a few comments on~\cite{gab}, where it has been argued that the field $F$ in the action (\ref{0action}) is nothing more than a Lagrange multiplier and can be consistently set to zero. This has the effect of eliminating the scalar ghost from the spectrum. However, it is important to realise what it {\it really} means to set $F=0$ in (\ref{0action}), from the point of view of the full bulk solution. To eliminate $F$ from (\ref{0action}), we need introduce a {\it non-normalisable} mode in $\phi(x, y)$ that cancels off the contribution from the bookkeeping term $h_\mn^{(F)}$. To see how this works, consider the general bulk solution for $\phi(x, y)$, retaining both normalisable and non-normalisable modes. This is given by
\be
\phi(x, y)=e^{-H|y|/2}\hat \phi(x)+e^{H|y|/2}\tilde \phi(x)
\ee
where the last term corresponds to the non-normalisable mode. In our analysis, we invoked the condition of normalisability to set $\tilde \phi(x)=0$, regardless of the value of the brane bending term $F$.  In order to set $F$ to zero in (\ref{0action}), as suggested in~\cite{gab}, we need to impose the boundary condition $\tilde \phi=-2F/H$. This boundary condition seems to violate a local, causal $4D$ description on the brane: if we change $F$ ever-so-slightly, we need the boundary condition for $\phi(x, y)$ as $y \to \infty$ to respond accordingly! Even if we accept this, there is strong evidence to suggest that retaining non-normalisable modes in the spectrum leads to further problems with ghosts. This was discussed in \cite{spec} in the context of the gravitational field of a  relativistic particle, or ``shockwave", on the brane. The non-normalisable modes contribute a {\it repulsive} potential on the brane, which would indicate the presence of ghosts.

It is also argued in \cite{gab} that in the presence of a heavy source, linearised perturbation theory breaks down below a Vainshtein radius $r_V$, and so one cannot make any conclusions as to whether or not the ghost is really there. Even in the region $r \gg r_V$ where  the linearised theory makes sense, it is not obvious that the linearised solution can always be smoothly continued inside $r_V$.  If this is indeed the case, one cannot use perturbations about the self-accelerating background (\ref{background}) to make any reliable cosmological predictions.  To proceed, we need to identify background solutions that take into account localised brane sources. To our knowledge, the only known exact solution with a localised brane source is  the shockwave~\cite{shock}. It would be interesting to study perturbation theory about this solution in order to see if the ghost remains.

\vskip .5in
\centerline{\bf Acknowledgements}
I would like to thank C. Charmousis, R.Gregory and N. Kaloper for a fruitful collaboration, aswell as the Universitat de Barcelona, where most of this work was completed.

\medskip

\end{document}